\documentclass[prx,twocolumn,showpacs,preprintnumbers,nofootinbib]{revtex4-1}
\usepackage{amsmath,bm,graphicx,hyperref,amssymb}

\usepackage{mathrsfs}
\usepackage{color}
\usepackage{yfonts}

\usepackage[authormarkup=none,
  draft
]{changes}
\definechangesauthor[name={YK},color=blue]{YK}

\def\LeaveRemarks{1}
\def\LeaveDeleted{1}

\setaddedmarkup{\uwave{#1}}
\setremarkmarkup{\,{\color{blue}\footnotesize$\langle\langle$#1: #2$\rangle\rangle$}\,}
\setdeletedmarkup{{\color{blue} \sout{#1}}}
\newcommand{\YK}[1]{\,{\color{cyan}\footnotesize$\langle\langle$YK: #1$\rangle\rangle$}\,}

\if\LeaveDeleted0
\setdeletedmarkup{}
\fi
\if\LeaveRemarks0
\setremarkmarkup{}
\renewcommand{\YK}[1]{}
\fi

\begin{document}

\title{Comprehensive study of the dynamics of a classical Kitaev spin liquid}

\author{A.~M.~Samarakoon$^{1,2}$}\email{ams4ux@virginia.edu}
\author{A.~Banerjee$^3$}
\author{S.-S.~Zhang$^4$}
\author{Y.~Kamiya$^5$}
\author{S.~E.~Nagler$^{3,6}$}
\author{D.~A.~Tennant$^7$}
\author{S.~-H~Lee$^2$}
\author{C.~D.~Batista$^{1,4}$}

\address{$^1$ Quantum Condensed Matter Division and Shull-Wollan Center, Oak Ridge National Laboratory, Oak Ridge, Tennessee 37831, USA.}
\address{$^2$ Department of Physics, University of Virginia, Charlottesville, Virginia 22904, USA .}
\address{$^3$ Quantum Condensed Matter Division, Oak Ridge National Laboratory, Oak Ridge, TN, 37830, USA. }
\address{$^4$ Department of Physics and Astronomy, University of Tennessee, Knoxville,
Tennessee 37996-1200, USA .}
\address{$^5$ Condensed Matter Theory Laboratory, RIKEN, Wako, Saitama 351-0198, Japan.}
\address{$^6$ Bredesen Center, University of Tennessee, Knoxville, TN, 37966, USA.}
\address{$^7$ Neutron Sciences Directorate, Oak Ridge National Laboratory, Oak Ridge, TN, 37830, USA. }

\date{June 2017}

\begin{abstract}
  We study the spin-$S$ Kitaev model in the classical ($S \to \infty$) limit  using Monte Carlo simulations combined with semi-classical  spin dynamics. We discuss differences and similarities in the  dynamical structure factors of the spin-$1/2$ and the classical Kitaev liquids. 
  Interestingly, the low-temperature and low-energy spectrum of the classical model exhibits a finite energy peak, which is the precursor of the one produced by the  Majorana modes of the $S=1/2$ model. The classical peak is spectrally narrowed compared to the quantum result, and can be explained by  magnon excitations within fluctuating one-dimensional manifolds (loops). Hence the difference from the classical limit to the quantum limit can be understood by the fractionalization of magnons propagating in one-dimensional manifolds. 
Moreover, we show that the momentum space distribution of the low-energy spectral weight of the $S=1/2$ model follows the momentum space distribution of zero modes of the classical model.
\end{abstract}

%75.10.Jm	Quantized spin models, including quantum spin frustration
%75.30.Ds	Spin waves
%03.65.Ge	Solutions of wave equations: bound states
%03.65.Nk	Scattering theory

\maketitle

\section{Introduction}

Quantum spin liquids (QSLs) have attracted great interest in both theoretical and experimental condensed matter physics due to their remarkable topological properties. Among many different proposals, the Kitaev model~\cite{Kitaev06} defined on the honeycomb lattice, is a prototypical two-dimensional (2D) QSL, which can be experimentally studied in iridium or ruthenium based materials~\cite{Jackeli09}. However, the lack of a symmetry-breaking order parameter poses a challenge for the experimental characterization of QSLs.
In the absence of a smoking-gun experiment, it is  important to characterize the  dynamical response of 
QSLs in order to identify signatures, which can guide the experimental search of these exotic states of matter~\cite{Barnejee16}.
The  computation of dynamical correlators of interacting \emph{quantum} spin systems in dimension higher than one is very challenging for state of the art techniques. For instance, the study of dynamics in the Kitaev-Heisenberg model, which is not integrable due to the additional Heisenberg interaction, was recently initiated by using a matrix-product state based $T = 0$ method ~\cite{Gohlke2017} and exact diagonalization~\cite{Gotfryd2017}. These $T=0$ techniques can only applied to relatively small clusters or quasi-one-dimensional lattice geometry.
Fortunately, the integrability of the pure Kitaev model allows for an exact  calculation of the magnetic structure factor, $S({\bf Q}, \omega)$, at $T=0$~\cite{Knolle14,Knolle15}
and for a controlled numerical calculation at any finite temperature $T$~\cite{yoshitake2016fractional,yoshitake17,yoshitake17b}.
This remarkable property is being used to identify proximates to Kitaev liquids~\cite{Barnejee16,banerjee2017neutron,do2017incarnation}. However, the actual model Hamiltonians of these materials are not integrable, so it is more challenging to assess
the effect of the additional Hamiltonian terms on $S({\bf Q}, \omega,T)$.

% and also by a low-energy effective field theory corresponding to a perturbed Kitaev liquid~\cite{Song2016}

\begin{figure}[!htbp]
 \includegraphics[width=0.55\textwidth,trim={6cm 4cm 3cm 3cm}]{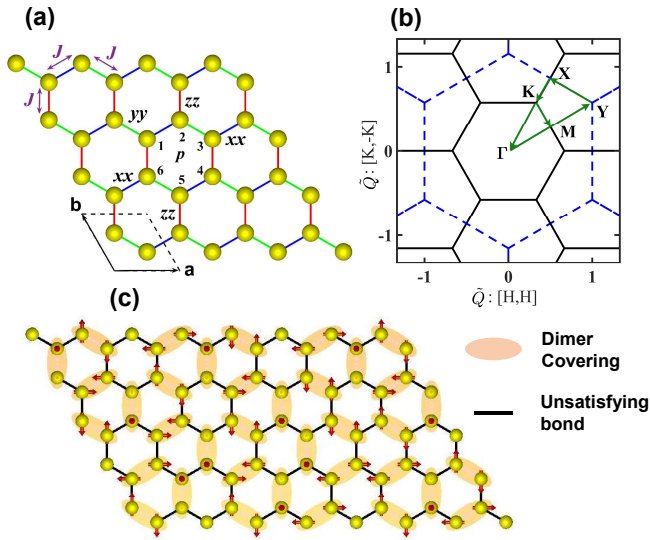}
 \caption{
   (a) Schematic illustration of  the Kitaev-Heisenberg model consisting of the Heisenberg interaction $J$ and the compass-like nearest-neighbor Ising interactions, with the associated spin component for each bond depending on the bond orientations ($xx$, $yy$, or $zz$).
The site indices around the plaquette $p$ correspond to the definition of the $Z_2$ flux operator $W_p$ in Eq.~\eqref{po}.    
(b) The first BZ (solid line) and the second BZ (dashed line)  of the honeycomb lattice.Here, $\tilde{Q} = Qa/2\pi$.
The green arrows indicate a path connecting high-symmetry points in the reciprocal space (i.e., K-$\Gamma$-M-Y-X-K-M), along which we evaluate $S(\mathbf{Q},\omega)$. (c) Example of a CN-ground state formed by antiferromagnetic dimer coverings.}
 \label{Fig1}
\end{figure}

Given the above considerations, it is relevant to ask if a  semi-classical  treatment can shed light on the dynamics of
the Kitaev QSLs. Semi-classical treatments are very useful for describing the low-temperature properties of unfrustrated magnets, whose low energy modes are quantized spin-waves or magnons.
  For instance, semiclassical dynamics simulations using an appropriate quantum-classical correspondence were found to produce a good description of the intermediate and high temperature regimes of the 2D $S = 5/2$ antiferromagnet Rb$_2$MnF$_4$, over all wavevector and energy scales, with a crossover temperature $\sim \theta_{CW}/S$ ($\theta_{CW}$ is the Curie-Weiss temperature)~\cite{Huberman07}.
It is clear, however, that the semi-classical  treatment cannot capture the intrinsically quantum mechanical nature of the low-energy excitations of quantum liquids.
 At first sight, this observation seems to render  semi-classical approaches  completely inadequate. Nevertheless,
  we will demonstrate that a semi-classical treatment of the Kitaev model can capture several properties of the dynamical structure factor of the 
  $S=1/2$ model, including a quite remarkable agreement above the quantun to classical crossover temperature $T_\mathrm{QC}$~\cite{Nasu14,Nasu15}.

The spin-$S$ Kitaev model with $S>1/2$ was introduced by Baskaran {\it et al.}~\cite{Baskaran08,baskaran2007exact} and it was subsequently studied by different groups~\cite{Price2012,Price2013,Ioannis17,chandra2010classical,nussinov2013compass}. This model is not exactly solvable, but it preserves the  $Z_2$ gauge structure of the $S=1/2$ model. The set of commuting  operators 
\begin{equation}
W_p= - \sigma^y_1 \sigma^z_2 \sigma^x_3 \sigma^y_4 \sigma^z_5 \sigma^x_6, 
\end{equation}
defined on each hexagonal plaquette of the honeycomb lattice (see Fig.~\ref{Fig1}), is generalized to 
\begin{equation}
W_p=  e^{i \pi (S^y_1 + S^z_2 + S^x_3 + S^y_4 + S^z_5 + S^x_6)}
\label{po}
\end{equation}
for arbitrary spin $S$. An immediate consequence of this local $Z_2$  symmetry is
that the two-spin correlator $\langle{S_i^\nu S_j^\nu}\rangle$ is nonzero only for $i = j$ and for the nearest-neighbor (NN) sites connected by a $\nu\nu$ bond ($\nu= x, y, z$)~\cite{Baskaran07,Baskaran08}. Consequently, both 
the quantum and the classical pure Kitaev models share the property of having a very short  correlation length $\xi \leq a$ ($a$ is the lattice space) for arbitrary temperature $T$. 
\begin{figure*}[!htp]
  \includegraphics[
    %width=\hsize,trim={3cm 3.5cm 3cm 3cm}
    width=\hsize,trim={0cm 5.5cm 0 5cm}
  ]{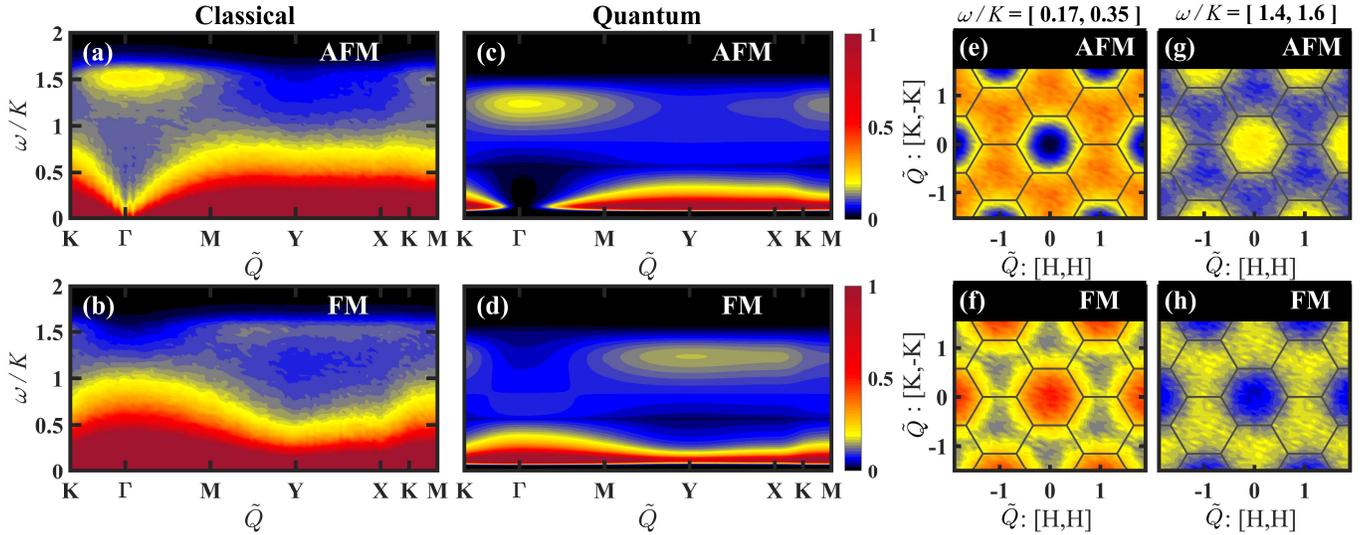}
  \caption{
Comparison of the dynamical structure factor, $S\left(\mathrm{Q},\omega \right)$,
in the classical limit $(S \to \infty)$ and  quantum limit ($S=1/2$)   of the pure Kitaev model ($J = 0$) at $T=0$.
Panels (a) and (b) show $S(\mathrm{Q},\omega )$ obtained from LL simulations of the classical  AFM and  FM Kitaev models, respectively. Panels (c)  and (d) show $S(\mathrm{Q},\omega )$ at $T=0$ for the $S=1/2$ Kitaev model  obtained in Ref.~\onlinecite{Knolle14}.   Constant energy cuts of  (e) AFM and (f) FM classical Kitaev liquids  obtained by integrating over the energy range $\omega /|K|=[0.17,0.35]$, corresponding to the low-frequency mode. Similar plots for the high-frequency mode are also shown for (g) AFM and (h) FM classical Kitaev liquids with the integration energy range $\omega /|K|=[1.4,1.6]$.
 }
  \label{Fig2}
\end{figure*}

As expected, the main differences between the quantum and the
classical limits of the model appear in the low energy sector. 
While the $S=1/2$ version of the model has  a
 unique QSL ground state, the ground state is massively degenerate in the classical limit.
The structure of the classical ground-state manifold corresponds to an exponentially large number 
of isolated points in the phase space, known as the Cartesian(CN)-ground states~\cite{Baskaran08}, as well as continuous families of intermediate states connecting one CN-ground state to another~\cite{Baskaran08,Ioannis17}.
The CN-ground states have each spin pointing along one of the three axes ($x$, $y$ or $z$), in such a way that one of the three bonds that arrive to a common site has the minimum possible energy, while the other two  have zero energy [see Fig~\ref{Fig1}(c)]. The zero modes associated with the continuous ground-sate degeneracy lead to a singular $\omega=0$  contribution to $S(\mathbf{Q}, \omega)$, which is naturally absent in the $S=1/2$ version of the model. Instead, 
the $S=1/2$ model leads to a structure factor $S(\mathbf{Q}, \omega, T=0)$, which  vanishes for $\omega \leq \Delta_v$, as a consequence of the finite activation gap $\Delta_v$ of the pair of bound $Z_2$ fluxes (or ``visons'') created by the application of a spin operator to the ground state.

Despite these qualitative differences between the low-energy magnetic response function of the quantum and
the classical Kitaev liquids at $T=0$, we will show in this manuscript that the magnetic structure factors, $S(\mathbf{Q}, \omega, T)$,  of both models  become very similar for $\omega$ and $T$ bigger than a quantum to classical crossover energy scale $T_{QC}$. Our results then suggest the possibility of
describing the thermally induced random-flux state at $T \gtrsim \Delta_v$ in the $S=1/2$ Kitaev model~\cite{Nasu14,Nasu15,Nasu17} with the classical liquid of the model obtained in the $S \to \infty$ limit. This observation can be exploited to identify  proximate quantum Kitaev liquid materials because $S(\mathbf{Q}, \omega, T)$ can be computed under control for any arbitrary deformation of the pure Kitaev Hamiltonian in the classical limit. 

\section{Zero Temperature Liquids}
The dynamics of the classical version of Kitaev model is studied by combining  Metropolis sampling and Landau-Lifshitz (LL) dynamics: 
\begin{equation}\label{eq:1}
{{d{\bm S}_i}\over {dt}}= {\bm S}_i \times {\bm B}_i 
\end{equation}
Where ${\bm B}_i$ is the effective local field (molecular field) acting on the spin ${\bm S}_i$. The temperature of the simulation is fixed during the Metropolis sampling and the Landau-Lifshitz dynamics starts from a randomly selected well-thermalized configuration. To study the effect of  relevant perturbations, which replace the Kitaev liquid by an ordered state at low-enough temperatures, we will consider the simple case of
the classical Kitaev-Heisenberg (KH) Hamiltonian~\cite{Price2012}\cite{Price2013}
with only nearest-neighbor (NN) interactions:
\begin{equation}
 H = K \sum_{\nu=\ x, y, z }{\sum_{{\left\{i,j\right\}}_\nu}{S^{\nu}_iS^{\nu}_j}}+J \sum_{\left\{i,j\right\}}{\mathbf{S}_i \cdot \mathbf{S}_j},
\label{KH}
\end{equation}
The index $\nu$ for the  variable ${i, j}$ indicates that the two n.n. sites $i$ and $j$ are connected by a $\nu\nu$ bond [see  Fig.~\ref{Fig1}~(a)]. To compare the results of this classical model against  the spin $S$ quantum version of the model, we normalize the classical spins
as $|\mathbf{S}_i|= \sqrt{S(S+1)}$.
The quantum version of this Hamiltonian has been proposed as a
model for iridium or ruthenium-based materials~\cite{Jackeli09,khaliullin2005orbital, chaloupka2010kitaev, chaloupka2013zigzag}. The magnetic structure factor $S(\mathbf{Q}, \omega, T)$ is obtained by Fourier transforming the real-space correlator $ \langle \mathbf{S}(\mathbf{r}_i, t) \cdot \mathbf{S}(\mathbf{r}_0, 0) \rangle$
evaluated from the LL dynamics over a finite period with periodic  boundary conditions.

We first focus on the pure Kitaev limit  ($J=0$). Figure~\ref{Fig2} shows the  magnetic structure factor $S(\mathbf{Q},\omega )$ of the classical and the quantum models at $T=0$ for both antiferromagnetic (AFM) and ferromagnetic (FM) cases. $S\left({\mathbf{Q}},\omega \right)$ is plotted along the BZ path (K$\Gamma$MYXK) shown in  Fig.~\ref{Fig1}(b). The calculations in the classical limit (CL),
shown in Figs.~\ref{Fig2}(a) and \ref{Fig2}(b), are averages over 120 LL simulations on a supercell of $20 \times 20$ unit cells (800 spins). The quantum limit (QL) calculations, shown in Figs.~\ref{Fig2}(c) and \ref{Fig2}(d), correspond to the exact result in the thermodynamic limit~\cite{Knolle14}. Remarkably, both the classical and the quantum Kitaev liquids are found to have two different almost
dispersionless modes centered at high and low frequencies ($\omega$) with striking similarities.

The high-energy mode is centered around the ${\boldsymbol \Gamma}$ ($\mathbf{Y}$) point for 
$K>0$ ($K<0$) and it is accompanied by a suppression of the low-energy spectral weight centered around the same wave vector.
This behavior is better illustrated by the contour plots shown in Figs.~\ref{Fig2}(e)--\ref{Fig2}(h).
These panels are constant frequency cuts of $S({\mathbf{Q}}, \omega)$, which show the distribution of spectral weight over momentum space. Figures~\ref{Fig2}(e) and \ref{Fig2}(f) correspond to the distribution of low-frequency modes (integral of $S(\mathbf{Q}, \omega)$ over the interval $\omega/|K| = [0.17, 0.35]$), while Figs.~\ref{Fig2}(g) and \ref{Fig2}(h) show 
the distribution of high-frequency modes (integral of $S(\mathbf{Q}, \omega)$ over the interval $\omega/|K| = [1.4, 1.6]$). 
As it is clear from these panels, the low-energy spectral weight is suppressed in the same region in momentum space where the distribution of high-energy spectral weight has a peak. This is the center of the first BZ for $K>0$ and the center of the second BZ for $K<0$ [see Fig.~\ref{Fig1}(b)].

\begin{figure*}
  \includegraphics[
    %width=\hsize,trim={3cm 3.5cm 3cm 3cm}
    width=\hsize,trim={3.5cm 4.2cm 2.5cm 4cm}
  ]{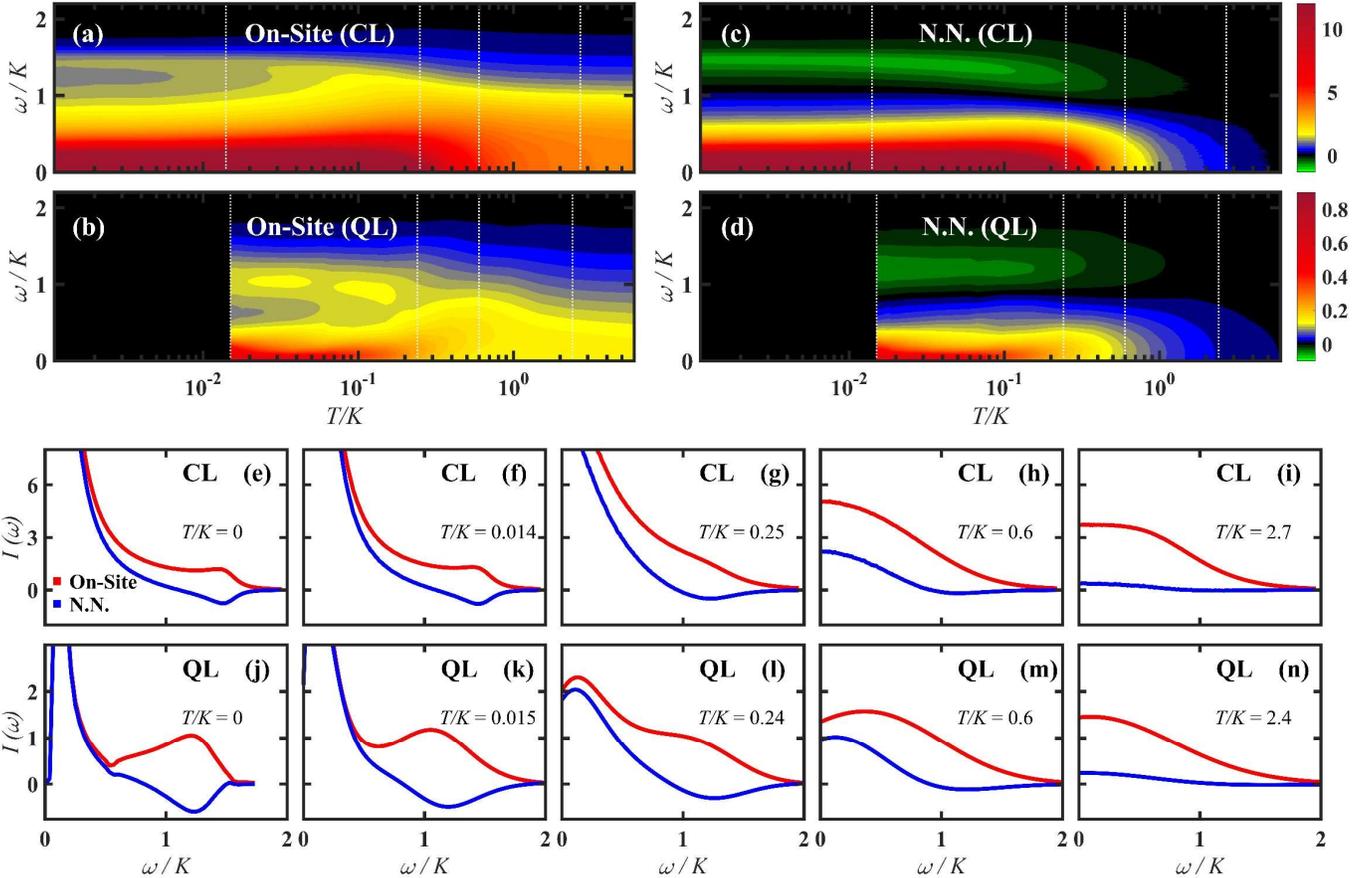}
  \caption{
 Temperature dependences of the dynamical spin structure factor decomposed into the on-site contribution, $S_0(\omega)$, and the NN contribution, $-S_1(\omega)$, for the classical Kitaev model and also for the $S=1/2$ Kitaev model:
      $S_0(\omega)$ for (a) $S = \infty$ and (b) $S = 1/2$ and $-S_1(\omega)$ for (c) $S = \infty$ and (d) $S = 1/2$.
Frequency-dependence of $S_0(\omega)$ and $-S_1(\omega)$ in the classical model at selected temperatures (e) $T/K = 0$, (f) 0.014, (g) 0.25, (h) 0.6  and (i) 2.7 [indicated by the dashed lines in panels (a) and (c)]. Similar plots for the $S = 1/2$ model are shown for comparison for (j)  $T/K=$ 0, (k) 0.015, (m) 0.24,  (l) 0.6 and (n) 2.4 [indicated by the dashed lines in panels (c) and (d)].
Note that the n.n. dynamical correlation function is $\pm S_1(\omega)$ for $K = \pm |K|$, while the on-site correlator is the same for both signs of $K$.
 }
 \label{Fig3}
\end{figure*}

To understand the differences and similarities between the classical and the quantum limits of the Kitaev model, it is instructive to go back to the real-space. Figure~\ref{Fig3} shows the real space spin-spin correlators for the classical and the quantum limits of the AFM model.
Figures~\ref{Fig3}(a) and \ref{Fig3}(b) include the on-site correlator for the CL and the QL, respectively. Similarly,
Figs.~\ref{Fig3}(c) and \ref{Fig3}(d) contain the NN correlator for the CL and the QL, respectively.
As we mentioned before, the local gauge structure shared by the quantum and the classical models leads to a real space spin-spin correlator
 that vanishes beyond NN sites. This implies that the spin structure factor in the pure Kitaev model for arbitrary $S$ can be decomposed as
\begin{equation}
    S^{\nu \nu}(\mathbf{Q},\omega) = S_{0}(\omega) +\cos(\mathbf{Q}\cdot\mathbf{u}_{\nu}) S_{1}(\omega),
  \label{eq:Sqw}
\end{equation}
  where $\nu=x,y,z$ and $\mathbf{u}_{\nu}$ is the relative vector between two NN sites connected by a $\nu\nu$ bond; $S_{0}(\omega)$ and $S_{1}(\omega)$ are the Fourier transformations into the frequency domain for the on-site and the NN dynamical spin correlators, respectively. This peculiarity leads to the sinusoidal $\mathbf{Q}$-modulation in the high- and low-energy peak intensities as illustrated in Figs.~\ref{Fig2}(e)--\ref{Fig2}(h). In other words, the similar wave vector dependence of the different modes in the classical and
the quantum limits are a direct consequence of the similar real space correlations shown in Fig.~\ref{Fig3}.

The real space correlators also exhibit a low  and a high-frequency peak in both models.  The low-frequency peak of the
$S=1/2$ model appears right above the small activation gap $\Delta_v$ for the pair of excited $Z_2$
gauge fluxes (visons) created by the action of a spin operator on the ground state.
The extended linewidths of the low and the high-energy peaks arise from the continuum of Majorana fermion excitations, which leads to a rather narrow low-energy peak and a broad high-energy peak.
In contrast, the low energy peak of the classical model extends down to zero frequency because the fluxes become gapless in this limit.  The CN-ground states are not eigenstates of the $W_p$ operators, implying that the classical limit $(S \to \infty)$ corresponds to a flux condensation (the flux number is no longer a good quantum number).
It is interesting to note that quadratic quantum fluctuations partially restore this quantum number by selecting the  CN-ground states, which maximize the number of hexagonal plaquettes with well-defined flux equal to zero (eigenvalue  of  $W_p$ equal to one).
These are the $3\times 2^{N/3}$ states with every single loop being an elementary hexagon~\cite{Baskaran08}.
The extensive degeneracy of the classical limit leads to  sharp $\delta$-function singularity in the spectral weight at $\omega=0$. The simple analysis that we present below explains the origin of this singularity
and of the high-energy peak centered around the $\Gamma$ point of the AFM model [see Fig.~\ref{Fig2}(a)].

\begin{figure}
 \includegraphics[
 %trim={0.5cm 2.0cm 0.5cm 2cm},clip,width=0.45\textwidth 
 width=\hsize,trim={0.5cm 6cm 0.5cm 5cm}
 ]{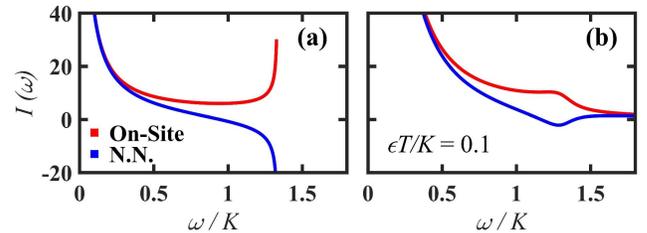}
 \caption{
 (a) Frequency dependence of $S_{0}(\omega)$ and $-S_{1}(\omega)$  associated with the CN-ground states in the classical Kitaev model at $T = 0$.  (b) Refined evaluations of $S_{0}(\omega)$ and $-S_{1}(\omega)$ with the inclusion of an artificial broadening $\epsilon=0.1K$, which mimics the effect of the continuum of the intermediate ground states connecting
 different CN-ground states through the slide transformations (see the text). }
 \label{Fig3p5}
\end{figure}

The main qualitative aspects of the dynamical structure factor of the classical model at $T=0$ can be captured by  the CN-ground states.  As explained in Ref.~\cite{Baskaran08}, the CN-ground states can be mapped into the close-packed dimer coverings of the honeycomb lattice by assigning one dimer to each ``satisfied'' bond (i.e., with its local energy taking the minimum value) [see Fig.~\ref{Fig1}(c)]. Within linear spin-wave theory~\cite{Baskaran08}, magnons for CN-ground states can only propagate along the one-dimensional paths of empty bonds because of the Ising nature of the interactions. These 1D paths become self-avoiding loops if we adopt closed boundary conditions, which fully cover the whole lattice (every spin site is visited by one and only one loop). The spin-wave Hamiltonian for each loop is invariant under translations by two sites along the loop~\cite{Baskaran08}. In other words, the unit cell of the loops has two sites, implying that each loop has two branches of magnetic modes: a flat branch of zero modes, $E_0(k)=0$, and second branch with a dispersion relation
\begin{equation}
  E (k)=2 |K| S \cos(k/2),
\end{equation}
with  $k$ being the momentum associated with the two-unit translation within a loop.

The top of the single-magnon band is at $k=0$, implying that the density of single-magnon states has a Van Hove singularity at $k=0$ for infinitely long loops. 
For the AFM model ($K>0$), the $k=0$ magnon wave function has the same phase for both sites in the unit cell.
Consequently,  the singular density of states  leads to a high-energy peak centered at the $\Gamma$ point [Fig.~\ref{Fig2}(a)]. On the other hand, the flat band of zero modes leads to a delta-like contribution at $\omega=0$. While the real space dynamical structure factors are obtained by averaging over all the CN-ground states, this average is dominated by loops of very long length because of the critical nature of the close-packed dimer coverings of the honeycomb lattice. 
(The fully-packed self-avoiding loops on the honeycomb lattice is a critical system that can host loops of infinite length~\cite{Blote1994,Kondev1996}.)
Consequently, we can approximate the average over loops by the result that is obtained for an infintely long loop (see Appendix):
\begin{equation}
  S_{0}(\omega)  \simeq  S \pi\frac{\bar{\rho} \left (\frac{\omega}{KS} \right )}{\omega}+ \frac{\pi S}{2} \delta(\omega)\int_{-\pi}^{\pi}\frac{dk}{\cos(\frac{k}{2})},
\label{s0}
\end{equation}
for the average on-site spin-spin correlator, and
\begin{equation}
 \theta S_{1}(\omega) \simeq  \left[\frac{\pi\omega S}{2 (KS)^2}- \frac{ \pi S} {\omega}\right]\bar{\rho} \left ( \frac{\omega}{KS} \right)-
 \frac{\pi S}{2} \delta(\omega)\int_{-\pi}^{\pi}\frac{dk}{\cos(\frac{k}{2})}.
\label{s1}
\end{equation}
for the NN spin-spin correlator, where $\theta=K/|K|$ is the sign of $K$.
The function $\bar{\rho}(x)=K S \rho(x)$ is the dimensionless density of single magnon states,
\begin{equation}
  \bar{\rho}(x)=\frac{2}{\sqrt{1-\left(\frac{x}{2}\right)^{2}}}.
\end{equation}

The $1/\omega$ singularity that appears in the first term of Eqs.~\eqref{s0} and \eqref{s1} arises from the 1D nature of the problem at the linear spin-wave level. This singularity must then be regularized by higher order corrections in the $1/S$ expansion, which restore the 2D nature of the problem by connecting different loops. Nevertheless, we will see that the linear spin-wave contributions  \eqref{s0} and \eqref{s1} are already enough to understand the main features of the numerical results.
In particular, the $1/\omega$ tail explains the broad low-energy spectral weight of the numerical results shown in Figs.~\ref{Fig2}(a) and  (b).

The second term of Eqs.~\eqref{s0} and \eqref{s1} corresponds to the singular contribution from the flat band of zero modes in each loop.  We note that these singular contributions to the on-site  and the NN correlation functions differ only by a minus sign. Equation~\eqref{eq:Sqw} then
implies that the singular contribution from the zero modes vanishes exactly at the $\Gamma$ point for the AFM classical Kitaev model: $S^{\nu \nu}(\mathbf{\Gamma},\omega) = S_{0}(\omega)+ S_{1}(\omega)$. This fact remains true for any loop length. 
In other words, the distribution of zero modes over the Brillouin zone (BZ) has a node at the $\mathbf{\Gamma}$  point for $K>0$. 
Moreover,
the singular contributions from the first terms of Eqs.~\eqref{s0} and \eqref{s1} (infrared singularity associated with the 1D nature of the loops)  also cancel exactly at the $\Gamma$ point.
The numerical result for this distribution is shown in Fig.~\ref{Fig2}(e) for $K>0$ and in \ref{Fig2}(f) for $K<0$.
 As expected from our analysis, $S(\mathbf{Q}, \omega \to 0)$ is suppressed around the $\mathbf{\Gamma}$ point for $K>0$. This ``hole" in the density of zero modes is a signature of the AFM {\it classical Kitaev liquid}. Similarly, the FM Kitaev liquid is characterized by a suppression of the density of zero modes around the $\mathbf{Y}$ point (center of the second BZ). In this case, the zero mode contribution cancels exactly at 
the $\mathbf{Y}$ point only for the component $S^{\nu \nu}(\mathbf{Y},\omega) = S_{0}(\omega) - S_{1}(\omega)$
with the $\nu\nu$ bond is parallel to $\mathbf{Y}$. Consequently, the singular weight contribution at $\omega \to 0$ is suppressed at
the $\mathbf{Y}$ point, but it does not vanish.
 In both cases, the missing low-energy spectral weight is shifted to the high-energy peak at $\omega \gtrsim |K|$, as it is shown in Figs.~\ref{Fig2}(g) and \ref{Fig2}(h).

As we anticipated, the zero modes are removed in the quantum limit because the massive ground state degeneracy is lifted by quantum fluctuations.  The net result is that the divergent spectral weight at $\omega=0$ is transferred to a small, but finite frequency region $\omega \gtrsim \Delta_v$. The lack of zero modes at the $\Gamma$ point suggests that the ``hole" in the low-energy spectral weight should still be present in the spectral weight distribution right above the two vison gap $\Delta_v$. This expectation is confirmed by the numerical results shown in Fig.~\ref{Fig4}~(a) for $K>0$ and Fig.~\ref{Fig4}~(e) for $K<0$. Consequently, {\it this suppression  of the low energy spectral weight and the associated shift to high energy is also a characteristic of the quantum Kitaev liquid}.
We note that low-energy spectral weight around the $\Gamma$ point can in principle be induced by perturbations that break
the two-flux selection rule (spin operators connecting subspaces that differ by two gauge fluxes). However, as it was shown
in Ref.~\onlinecite{Song2016}, the rule is broken to fourth order in the typical perturbations of the Kitaev Hamiltonian, implying that the low-frequency spectral should remain very small.

Figure~\ref{Fig3p5}(a) shows the on-site, $ S_{0}(\omega)$, and the NN, $ S_{1}(\omega) $, dynamical structure factors given by Eqs.~\eqref{s0} and \eqref{s1}, respectively. These quantities are only an approximation of the exact $S(\mathbf{Q}, \omega, T=0)$  because the average is taken over the CN-ground states. The missing ground states correspond to the continuous deformations that connect different CN-ground states~\cite{Baskaran08}.

Let us consider an  intermediate ground state, which is obtained by continuous ``slide'' transformations~\cite{Baskaran08}
of a given CN-ground state. A slide transformation only involves spins along either a closed loop or an infinitely long string corresponding to alternating dimer and empty bonds. For small transformations, magnons
will still propagate mainly along the loops of the ``parent" CN-state. However, the corresponding 1D spin-wave Hamiltonian  is no longer translationally invariant. Moreover, magnons can tunnel between different loops because the spins on the ``satisfied bonds" are
no longer parallel to the Ising anisotropy axis of the bond.
In other words, the spin-wave Hamiltonian of the intermediate state  can be regarded as 
a disordered version of the spin-wave Hamiltonian of the parent CN-state. To zeroth order, the effect of disorder is to broaden the quasi-particle  peaks of the parent CN-state. Figure~\ref{Fig3p5}(b) shows $S_{0}(\omega)$ and $S_{1}(\omega)$ after introducing an effective broadening $\epsilon=0.1K$.
These curves reproduce quite well the numerical results shown in Fig.~\ref{Fig3}(e). In particular, the effective disorder introduced by the ``intermediate" ground states broadens the high-frequency peak originated by the Van Hove singularity in the density of states.
The success of  such a minimal perturbative treatment of intermediate states relies on the fact that  CN-ground states maximize the number of zero modes, i.e., most of the classical ground states are small deformations of  CN-ground states~\cite{Baskaran08}. Although this property does not lead to an order-by-disorder phenomenon at any temperature, it may renormalize the effective stiffness   leading to a nontrivial $T$-dependent power law in the short-range decay of the energy density correlator~\cite{Lin2014}.

Finally,  1D magnons must decay into pairs of spinons  upon inclusion of quantum fluctuations beyond linear spin-wave theory. The resulting two-spinon continuum can be regarded as a precursor of the Majorana modes, which appear in the 
$S=1/2$ limit. The main effect of the two-spinon continuum is to broaden the high-frequency peak, in agreement with the result obtained for the quantum limit of the model [see Fig.~\ref{Fig3}(b), (d) and (j)].

 Thus, to summarize our discussion on the low-$T$ classical liquid state, the high-energy peak of $S({\bm Q},\omega)$ is common to  the quantum ($S=1/2$) and the classical  ($S \to \infty$) $T=0$ liquids. The classical limit of the model provides a new insight for understanding the origin of this peak, which has been used as a fingerprint of the proximity to a Kitaev liquid state~\cite{Barnejee16,banerjee2017neutron,do2017incarnation}. The classical model also provides a new insight for understanding the momentum dependence of the low-energy spectral weight distribution of the $S=1/2$ model. This distribution  is very similar to the distribution of the $\omega=0$ spectral weight induced by the zero modes of the classical model. These facts establish a clear connection between the $T=0$ spectra of the classical and quantum liquids.
 %with the essential difference in whether the vison gap is zero or finite.

\section{Finite Temperature Liquids}
 
\begin{figure}[!htbp]
 \includegraphics[trim={7cm 4cm 0 3cm},clip,width=0.73\textwidth]{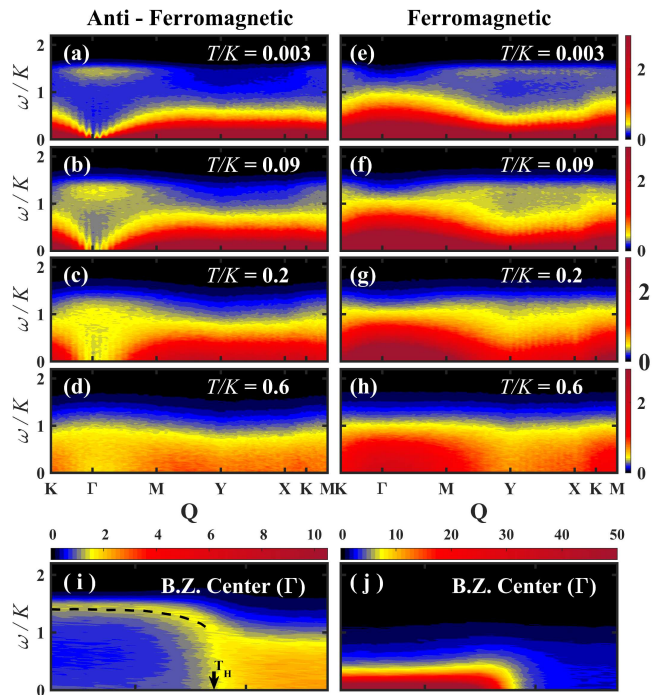}
 \caption{
 Temperature evolution of $S(\boldsymbol{\mathrm{Q}}, \omega )$ in the classical Kitaev model calculated along a BZ path shown in Fig.~\ref{Fig1}~(b) for the AFM coupling [(a)--(d)] and the FM coupling [(e)--(h)] at (a),(e) $T/K=$ 0.04, (b),(f) $T/K=$ 1.1, (c),(g) $T/K=$ 2.9 and (d),(h) $T/K=$ 7.5. Panels (i) and (j) show the temperature dependence  of the intensity at the  $\Gamma$-point for the AFM and the FM models, respectively. The dashed line in (i) traces the peak position of the high energy mode.}
 \label{Fig4}
\end{figure}

The $S=1/2$ Kitaev model can be mapped into a gas of free Majorana fermions interacting with static $Z_2$ gauge fields~\cite{Kitaev06}.  Because of the quadratic nature of the action, the fermionic degrees of freedom can be integrated out to obtain an effective classical action for the $Z_2$ variables, which can simulated by using Monte Carlo simulations~\cite{Nasu14,Nasu15}, similar to other problems of noninteracting fermions coupled to classical degrees of freedom~\cite{Weisse2006}. To evaluate the dynamical spin structure factor at finite temperatures, one needs to combine such Monte Carlo samplings with a quantum Monte Carlo solver~\cite{yoshitake17b}.
To compare against the  results for  $S(\mathbf{Q},\omega, T)$ of the $S=1/2$ Kitaev model [see Fig.~\ref{Fig3}], extracted from Refs.~\cite{yoshitake2016fractional} and \cite{yoshitake17}, we present the corresponding results for 
the classical Kitaev model in Figs~\ref{Fig3} and \ref{Fig4}.

As shown in Figs.~\ref{Fig4}(d) and \ref{Fig4}(h), the high-temperature paramagnetic (PM) state of both the  AFM and FM 
classical 
Kitaev models exhibits a characteristic broad diffusive peak with a small $\mathbf{Q}$-dependence.
Figures \ref{Fig4}(c), \ref{Fig4}(g), \ref{Fig4}(b), and \ref{Fig4}(f), show that the
low-frequency diffusive mode becomes more structured upon  decreasing temperature  and an additional mode emerges at the  BZ center for $K>0$ and at the center of the second BZ ($\mathbf{Y}$ point) for $K<0$. Finally, at very low temperatures [see Figs.~\ref{Fig4}(a) and \ref{Fig4}(e)] the upper mode goes up in energy and it  separates  from the low-frequency spectral weight. 

Figures~\ref{Fig4}(i) and \ref{Fig4}(j)
show the temperature dependence of $S(\Gamma, \omega)$ for the AFM and the FM cases, respectively. The 
dashed line in Fig.~\ref{Fig4}(i) indicates the temperature evolution of the high-frequency peak.   This peak merges with 
the low-energy peak at a temperature scale $T_H \sim 0.3-0.4K$, which  roughly
coincides with the high-temperature peak of the specific heat curve of the quantum ($S=1/2$) version of the model~\cite{yoshitake2016fractional}.
$T_H$ is also the temperature at which the high-energy mode of the $S=1/2$ Kitaev model merges with the low-energy mode~\cite{yoshitake2016fractional,yoshitake17}. 

The quantum to classical crossover occurs at the temperature scale $T_{QC} \simeq \Delta_v$, which is significantly lower than $T_H$. Degenerate Hamiltonian eigenstates whose number of fluxes differ by two only exist for energies bigger than $\Delta_v$, as it is indicated by the finite value  of $S_0(\omega=0)$ for the $S=1/2$ Kitaev model [see Fig.~\ref{Fig3}(l)]. Linear combinations of these degenerate states produce  eigenstates with non-zero $\langle {\bm S}_j \rangle$ (the total flux is no longer a good quantum number), which can be regarded as ``classical states".

A large concentration of $Z_2$ fluxes is induced at  $T > T_{QC} \simeq \Delta_v$. These fluxes act as scattering centers for the Majorana fermions, which loose their coherence when the distance between scattering centers becomes comparable to their wave-length. This condition is fulfilled at $T > T_{QC}$ because the average distance between thermally activated fluxes becomes of order one lattice space. Figures~\ref{Fig3}(g) and (l) show that the dynamical structure factors of the quantum and classical AFM models are very similar for  $T/K=0.25$ and $\omega > \Delta_v$. The quantum character of the liquid is manifested at low temperatures in the low-frequency dip at $\omega < \Delta_v$ visible even up to T/K = 0.6 K (Fig. ~\ref{Fig3}(m)). The resulting low-frequency peak is then a remnant of the 2-vison gap.

%The quantum nature of the model is manifested by the low-frequency peak at $\omega \simeq \Delta_v$.

\begin{figure}[!htbp]
 \includegraphics[
 trim={5.0cm 4.0cm 4.5cm 4.0cm},clip,width=0.53\textwidth
 %width=\hsize,trim={5.0cm 4.0cm 4.5cm 4.0cm} 
 ]{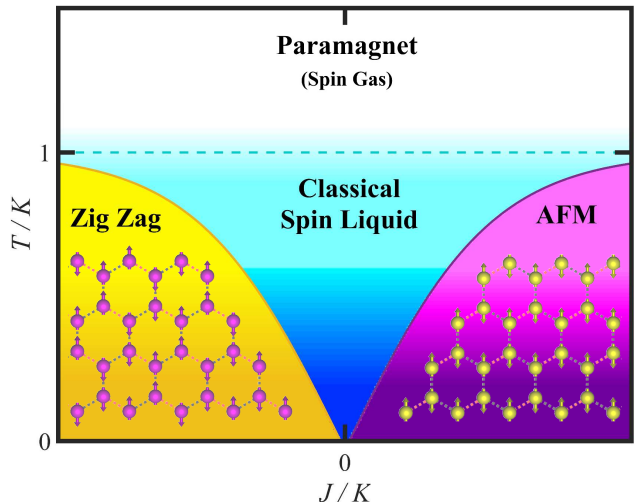}
 \caption{
 Schematic phase diagram of  the classical K-H model with AFM ($K > 0$) Kitaev interaction. The color gradient  denotes the  quantum-classical crossover region. The solid lines represent phase transitions. We note that the quantum spin liquid phase is stable over a finite interval
   of $J/K$ values around $J=0$ in the quantum limit ($S=1/2$).}
 \label{Fig5}
\end{figure}

\section{Kitaev-Heisenberg Model}

\begin{figure}[!htbp]
 \includegraphics[
  trim={7.8cm 3.2cm 7cm 3.0cm} ,clip,width=0.53\textwidth
 ]{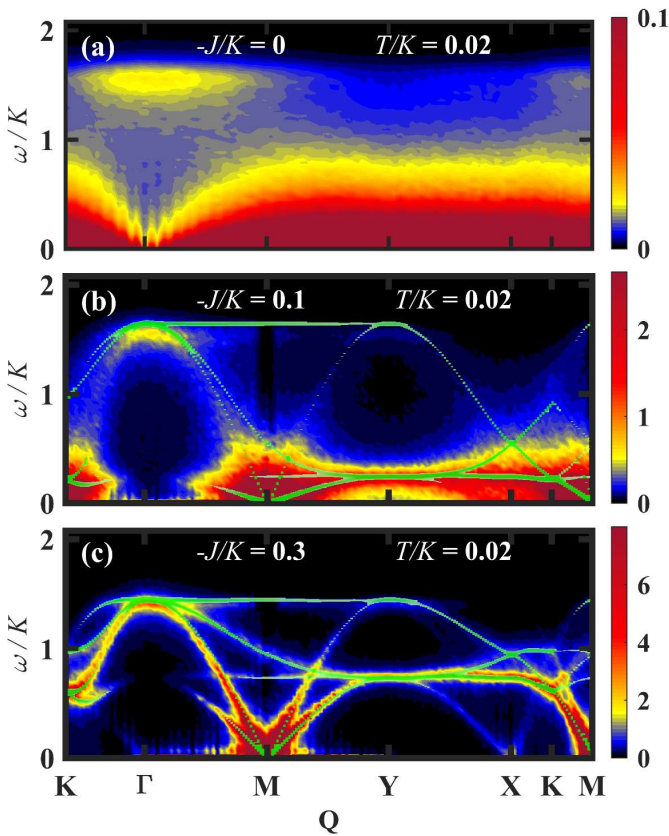}
 \caption{
   $S(\mathbf{Q}, \omega)$ for the KH model with  FM Heisenberg exchange ($J<0$)  and AFM Kitaev interaction $(K>0)$ as a function of  $-J/K$. The green lines correspond to the $S(\boldsymbol{\mathrm{Q}}, \omega )$ obtained from linear spin-wave theory 
 (the line thickness indicates the intensity ).
 Panel (a) corresponds to the pure Kitaev model ($J=0$), while panels (b) and (c)  correspond to $J/K=-0.1$ and $J/K=-0.3$, respectively. Panels (c) shows that the magnon modes are more sharply defined away from the Kitaev point. Panel  (b) clearly shows that the magnon modes become less defined upon approaching the Kitaev point because of the increasing importance of non-linear effects captured by the LL simulation.}
 \label{Fig6}
\end{figure}

A big advantage of the classical limit of the model is that we can study the evolution of $S(\mathbf{Q}, \omega )$
away from the Kitaev point. In contrast to the quantum case, an arbitrarily small perturbation is enough to replace the $T=0$ liquid with
a magnetically ordered phase that is also stable  at finite temperatures.
This ordered phase has three different regimes: a low temperature regime, $ T \ll T_N$, in which the magnetic structure factor is dominated by the single-particle excitations of the ordered state (spin-wave dispersion),  an intermediate temperature regime, $T_N < T \lesssim K$, with liquid-like correlations, and a high-temperature paramagnetic state, $T \gg K$, that can be regarded as a ``spin-gas" because the off-site correlations are negligibly small in comparison to the on-site correlations. 
Price and Perkins~\cite{Price2012}\cite{Price2013} argued that the ordering occurs via two  consecutive Berezinskii-Kosterlitz-Thouless transitions, as in the $C_6$ clock model, with a small critical phase in between.

The  KH model  \eqref{KH} is one of the simplest Hamiltonians that can be used to study the three regimes
(in what follows, our discussion excludes details related to critical phenomena at $T \approx T_N$).
Given that the intermediate spin-liquid regime only exists in the proximity of a $T=0$ liquid phase (see Fig.~\ref{Fig5}), 
this regime can be used to detect proximate quantum spin liquid behavior. If $T_N > T_{QC}$, the liquid-like regime is described by the classical limit of the model, implying that finite-$T$ classical spin dynamics can be exploited to identify magnets near a $T=0$ quantum melting point.

As shown in Fig.~\ref{Fig5}, the ground state of the classical KH model has zig-zag ordering for  $J/K <0$ and a two-sublattice AFM ordering for $J/K >0$  ($K>0$).   At $T=0$, the spin-wave dispersion  deviates  from the {\it linear} spin-wave spectrum (green lines in
Fig.~\ref{Fig6})  upon approaching $|J|/K=0$. These deviations 
arise from the non-linear effects associated with spin fluctuations towards the large manifold of classical states that become ground states for $J=0$. We note that this non-linearity may have a different manifestation in the quantum  $S=1/2$ model. In particular, the magnon modes of the $S=1/2$ model should become weakly bounded pairs of Majorana fermions upon approaching the transition into the spin liquid phase
(here we are assuming that the transition is continuous or quasi-continuous). 
It is clear that this intrinsically quantum phenomenon cannot be captured by the classical limit of the model. 
However, as we discuss below, the classical model is still capable of capturing the  evolution of the high-energy features of the magnon spectrum.
Moreover, based on the results discussed in the previous sections, the classical model can describe  the second way of approaching the spin liquid regime, which is by increasing temperature at a fixed value of $J/K$.

Figure~\ref{Fig6} shows the evolution of the low-temperature $ S(\mathbf{Q}, \omega )$ as a function of $J/K$, from the pure AFM classical Kitaev model [Fig.~\ref{Fig6}(a)] to  $-J/K=0.3$ [Fig.~\ref{Fig6}(c)]. As expected, $ S(\mathbf{Q}, \omega )$  exhibits a sharply defined spin wave dispersion with a pseudo-Goldstone mode at the  M-point (ordering wave vector)  well inside the ZZ phase. We note that there are three inequivalent M-points corresponding to the possible directions of the FM ZZ chains.  Upon reducing $|J|/K$, the magnon modes become less defined due to the increasing relevance of non-linear effects triggered by the proximity to the ($J=0$) spin liquid point. In particular, the spectrum obtained for $J/K=-0.1$ [Fig.~\ref{Fig6}(b)]  shows an overall softening of the acoustic magnon modes, except for the region around the $\Gamma$ point where the spectral weight   remains around the original optical mode. The evolution of this ``high-energy" feature should be common to both the classical and quantum limits of the model.
In other words, this unusual behavior can be used to detect the proximity to a Kitaev quantum spin liquid, as long as the transition between the ZZ phase and the liquid state remains continuous (or quasi-continuous) in the $S=1/2$ limit.  

 \begin{figure*}[!htbp]
 \includegraphics[
trim={3.8cm 3.2cm 3.8cm 3.0cm} ,clip,width=0.9\textwidth 
 ]{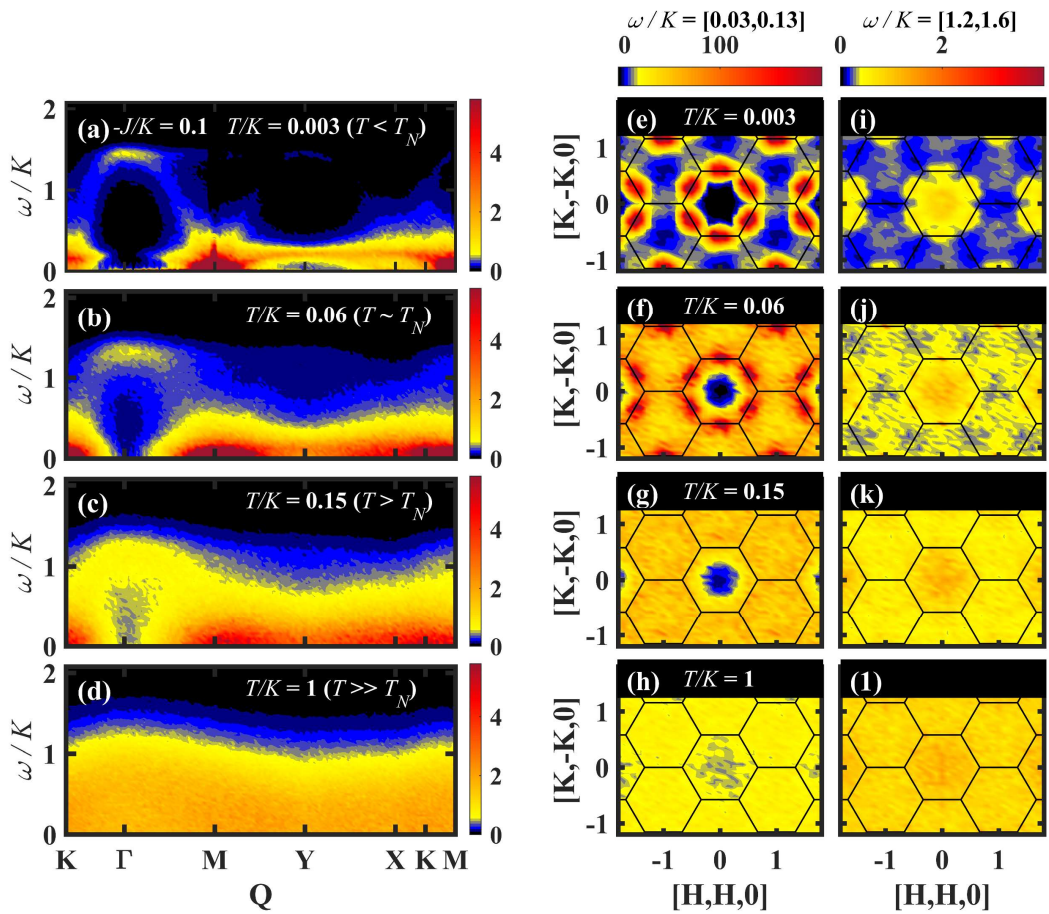}
 \caption{%
   $\ S(\mathbf{Q}, \omega )$ obtained for the KH model with  $J /K=-0.1$. The different panels show  $S(\boldsymbol{\mathrm{Q}}, \omega )$ along the BZ path K$\Gamma$MYXKM for (a) $T/K=0.02$ ($T \ll T_N$), (b) $T/K=0.09$ ($T \simeq T_N$), (c) $T/K=0.17$ ($T> T_N$) and $T/K=1.7$ ($T \gg T_N$). The panels on the right hand side show the distribution of  spectral weight over the BZ: the first column
shows the low-energy spectral weight ($0.03 < \omega < 0.13$), while the second column shows the spectral weight in the frequency interval of the high-energy peak  ($1.2 < \omega < 1.6$). 
\label{Fig7}
 }
\end{figure*}

As shown in Fig.~\ref{Fig5}, the spin liquid state can also be accessed by increasing $T$ at a fixed value of $J/K$. Fig.~\ref{Fig7} shows the temperature evolution of $\ S(\bf{Q}, \omega )$ for $J/K=-0.1$. The corresponding N\'{e}el temperature is $T_N \approx 0.06K$, implying that there is a temperature window, $|J| < T < K$, above $T_N$, where $S(\mathbf{ Q}, \omega, T )$ should be very similar to the dynamical structure factor of the pure Kitaev model ($T$ is high-enough to suppress the magnetic correlations induced by the Heisenberg interaction). This is true for any other small perturbation that can be added to the pure Kitaev model. 
Fig.~\ref{Fig7}~(c)  confirms this expectation: $ S(\mathbf{Q}, \omega)$ is very similar to the result shown in \ref{Fig4}~(c) for the pure Kitaev model ($J=0$). As shown in  Fig.~\ref{Fig7}~(g), the low frequency region shows the signature of a Kitaev liquid, with a rather uniform distribution of low energy modes over the BZ, which is suppressed around the ${\boldsymbol \Gamma}$ point.
Correspondingly, Fig.~\ref{Fig7}~(k) shows the opposite behavior for the distribution of high-energy modes over the BZ ($1.2 < \omega/K < 1.6$). 
The rest of the panels on the right hand side of Fig.~\ref{Fig7} show the continuous redistribution of spectral weight upon moving towards the low and high temperature regimes.
Given that quantum corrections are small above $T_{QC} \simeq \Delta_v \simeq 0.06K$,
the result obtained at $\omega > \Delta_v$ with the classical spin model represents the $S(\mathbf{Q}, \omega, T )$ of the QL (see Fig.~\ref{Fig3}). 
%Similarly, \ref{Fig4}~(c)  also shows the high-energy modes localized around the ${\boldsymbol \Gamma}$ point, which is another fingerprint of the AFM Kitaev liquid.

We remark that the Kitaev liquid state that appears in the intermediate temperature regime is independent of the particular model Hamiltonian, as long as the additional terms can be treated as small perturbations relative to the Kitaev contribution (separation of energy scales). 
This observation is relevant for  candidate materials based on $4d$ and $5d$ elements because their microscopic Hamiltonian models include multiple interaction terms, whose values are still 
uncertain~\cite{yadav2016kitaev,winter2016challenges,wang2016theoretical,kim2016crystal,kim2015kitaev,rousochatzakis2017classical,
sizyuk2016selection,kim2016quasimolecular,chaloupka2016magnetic,janssen2016honeycomb,hou2016unveiling, winter2017breakdown,ran2017spin,sela2014order}.

\section{Conclusions}

We have shown that the  dynamical spin structure factors of the classical and the quantum ($S=1/2$) limits of the Kitaev model become very similar above a crossover temperature $T_{QC} \simeq \Delta_v$.
Moreover, both structure factors exhibit similar qualitative behavior in their high-frequency response ($ \omega \gtrsim K$) even at $T=0$.  This ``high-energy" response is characterized by a broad peak above $\omega \simeq K$, which is centered around the ${\boldsymbol \Gamma}$ point for AFM Kitaev model ($K>0$) and around the $\mathbf{Y}$ point for the FM Kitaev model ($K<0$). Correspondingly, the singular spectral weight at $\omega=0$, produced by the zero modes of the classical model, is suppressed around the ${\boldsymbol \Gamma}$ point for AFM Kitaev model and around the $\mathbf{Y}$ point for the FM Kitaev model. This dip in the momentum space dependence of the low-energy spectral weight is still present in the $S=1/2$ model. The main difference is that the low-energy modes of the quantum $S=1/2$ Kitaev model appear right above the two-vison activation gap $\Delta_v$.

In the classical limit, the low-energy modes of the CN-ground states correspond to single-magnons states that propagate in 1D loops \cite{Baskaran08}. The high-energy peak of the classical Kitaev model arises from the singular density of single-magnon states at the top of the dispersive branch of excitations.  This Van Hove singularity is smoothed out by the deformations of the CN-ground states into the non-Cartesian ground states (valleys) that lead to the flat branch of zero modes and also by thermal fluctuations.
As discussed in Ref.~\onlinecite{Baskaran08}, the 1D magnons of the CN-ground states must decay into fractionalized excitations upon inclusion of quantum fluctuations. These excitations can be regarded as  precursors of the Majorana modes obtained in the quantum $S=1/2$ limit. In particular, this fractionalization leads to an additional broadening of the high-energy peak of $S(\mathbf{Q}, \omega)$, which explains why the peak of the classical model is narrower than the peak of the $S=1/2$ model.

Our results provide a systematic procedure for identifying proximate quantum spin liquid behavior of real materials. A dip in the density of low-energy modes at the ${\boldsymbol \Gamma}$ ($\mathbf{Y}$) point must be accompanied by a high-energy peak around the same wave-vector for 
$K>0$ ($K<0$). For materials that exhibit low-temperature magnetic ordering, the signatures of the Kitaev liquid should appear over an intermediate temperature window above the ordering temperature. As long as $T \gtrsim T_{QC}$, the classical approach can be used in this temperature window  to obtain a good approximation of $S(\mathbf{Q},\omega)$  for the $S=1/2$ model.

The analysis presented here can have more general implications for other quantum liquids with extensive ground state degeneracy in the classical limit.
Given the lack of magnetic ordering, one needs to find an alternative low-energy characterization of the liquid state. Our results suggest that the distribution of zero modes over the BZ provides clear signature of the classical liquid, which is inherited by the distribution of low-energy modes of the quantum spin liquid. Given that such a distribution can be measured with inelastic neutron scattering,\cite{choi2012spin,banerjee2017neutron,do2017incarnation} this experimental technique can play a crucial role in the characterization of quantum spin liquids. Moreover, the quantum to classical crossover  can be exploited for computing other dynamical correlation functions and transport properties of quite general quantum spin models at $T >T_{QC}$.

\begin{acknowledgments}
We thank Y. Motome for enlightening discussions and for providing the finite temperature results for the $S=1/2$ Kitaev model shown in 
Fig.~\ref{Fig3}. We additionally thank J. Knolle and R.Moessner for providing the $T=0$ results for the $S=1/2$ Kitaev model shown in 
Fig.~\ref{Fig2}. Work at ORNL is supported by the US DOE, Office of Science, Basic Energy Sciences, Scientific User Facilities Division.
Y.K.~acknowledges support by JSPS Grants-in-Aid for Scientific Research under Grant No.~JP16H02206.
C.D.B. and S.-S. Zhang acknowledge support from the Los Alamos National Laboratory LDRD program.
\end{acknowledgments}

\appendix

\section{Linear Spin Waves for CN-ground States}
  \label{SW-CN}

The ground state of the classical Kitaev model has an extensive degeneracy. The subset of CN-ground states can be mapped to the the close-packed dimer coverings of the honeycomb lattice.~\cite{Baskaran08} The empty bonds of each dimer covering form  self-avoiding (SW) paths, which are  loops for closed boundary conditions. Within linear spin-wave theory, magnons can only propagate along these 1D paths to lowest order in a $1/S$ expansion.  To compute the spin-wave Hamiltonian in each loop for a given CN-ground state, it is convenient to use a  twisted reference frame for the original Hamiltonian defined on a given loop, where the local $z$-axis on a given site is chosen to be parallel to the spin direction:
\begin{equation}
H_{1D}=K \sum_{i=1}^{m}( \tilde{S}_{i1}^x \tilde{S}_{i2}^x + \tilde{S}_{i2}^y \tilde{S}_{i+1,1}^y) -KS \sum_{i=1}^{m}\sum_{\alpha=1}^{2} \tilde{S}_{i\alpha}^z ,\label{eq:Hsw}
\end{equation}
where $m=n/2$ with $n$ being the number of sites on the loop. The local reference frame is chosen in such a way that  two adjacent sites have the same local $x$ ($y$)
axis if they are connected by a $xx$ ($yy$) bond. With this construction, the closed boundary condition can be periodic or anti-periodic depending on the direction of the last spin ~\cite{Baskaran08}. In this reference frame, the Hamiltonian is invariant under translations by two lattice sites. Correspondingly, the index $\alpha=1,2$ denotes the two sites on the effective unit cell.
The second term of Eq.~(\ref{eq:Hsw}) represents an effective perpendicular magnetic field generated by the adjacent 1D path through the antiferromagnetic
interaction on the dimer.

After a  Holstein-Primakoff transformation, 
\begin{eqnarray}
\tilde{S}^z_{i\alpha} &=& S - a_{i\alpha}^{\dagger} a_{i \alpha}, \\
\tilde{S}^x_{i\alpha} &=& \frac{1}{\sqrt{2}}(a_{i\alpha}^{\dagger}+a_{i\alpha}),\;\tilde{S}^y_{i\alpha}=\frac{1}{\sqrt{2}i}(a_{i\alpha}^{\dagger}-a_{i\alpha}).
\end{eqnarray}
the spin wave Hamiltonian (\ref{eq:Hsw}) can be rewritten as
\begin{align}\label{eq:Hsw2}
H_{sw} & =\frac{KS}{2}\sum_{i=1}^{m}\sum_{\alpha=1}^{2}\left(a_{i\alpha}^{\dagger}a_{i\alpha}+a_{i\alpha}a_{i\alpha}^{\dagger}\right) \nonumber \\
 & +\frac{KS}{2}\sum_{i=1}^{m}\left(a_{i1}^{\dagger}a_{i2}^{\dagger} + a_{i1}^{\dagger}a_{i2} + h.c. \right) \nonumber \\
 & -\frac{KS}{2}\sum_{i=1}^{m}\left(a_{i,2}^{\dagger}a_{i+1,1}^{\dagger} - a_{i,2}^{\dagger}a_{i+1,1} +h.c. \right).
\end{align}
Given the translational
symmetry of $H_{sw}$, we can diagonalize it by  Fourier transforming the creation and annihilation operators:
\begin{equation}
a_{i\alpha}=\frac{1}{\sqrt{m}}\sum_{k}a_{k\alpha}e^{ik(i+\delta_{\alpha})},
\end{equation}
where $\delta_{\alpha}$ refers to the displacement within each unit cell. 
After Fourier transforming and applying a Bogoliubov transformation to  $H_{sw}$, we
obtain the diagonal form
\begin{eqnarray*}
H_{sw} & = & \sum_{k} E(k) \beta_{k}^{\dagger}\beta_{k}+ E_0 \sum_{k} \gamma_{k}^{\dagger}\gamma_{k}+ K S \sum_{k}\rvert\cos(\frac{k}{2})\rvert,
\end{eqnarray*}
where $E_0=0$. The branch of zero modes arises from the continuum of (non-CN) ground states  connecting different CN-ground states. The dispersion relation of the dispersive branch is~\citep{Baskaran08}
\begin{equation}
E(k)  =2 |K| S \cos(k/2).
\end{equation}

\subsection{Dynamical structure factor}

To the lowest non-trivial order in the $1/S$ expansion, the dynamical structure factor in momentum-frequency space for a given ground state, $|0 \rangle$, has only contributions from the transverse spin components in the local reference frame
\begin{align}
\tilde{S}_{\alpha\beta}^{\mu\nu}(k,\omega) =4\pi^{2} \!\!\!\sum_{n=1,2}  \!\!\! \langle 0 \rvert\tilde{S}_{k,\alpha}^{\mu}\rvert k,n\rangle\langle k,n\rvert\tilde{S}_{-k,\beta}^{\nu}\rvert 0 \rangle\delta(\omega-E_{q,n}).
\end{align}
As we explained in Sec.~I, the local gauge structure of the Kitaev Hamiltonian, $[ H_K, W_p]=0$, implies that the real space spin-spin correlators must vanish for distances bigger than one lattice parameter. Based on that observation, we will only compute the on-site and the NN spin-spin correlators that arise from 
taking the average over all the CN ground states. Note that a more rigorous calculation of the $T=0$ spin-spin correlator  should also include the non-CN ground states. However, a calculation based on the just the CN-ground states is enough to capture the main qualitative features of the dynamical structure factor obtained from our numerical simulations of the classical AFM Kitaev model. 

Finally, given the critical nature of the dimer coverings of the honeycomb lattice, the loop length has a power law distribution, implying that most of the loops containing a given site (for the on-site correlator) and a pair of sites (for the two-site correlator) have a very long length. Consequently, we will assume that the average over CN states is dominated by the result for infinitely long  loop length. 

\subsection{On-site dynamical structure factor}

The on-site dynamical structure factor is obtained by averaging over both sites of the unit cell of the loop:
\begin{align}
\tilde{S}_{0}^{xx}(\omega) =\frac{1}{2m}\sum_{k}\left[\tilde{\chi}_{11}^{xx}(k,\omega)+\tilde{\chi}_{22}^{xx}(k,\omega)\right],
\end{align}
Replacing the creation and annihilation operators of Holstein-Primakoff bosons with
Bogoliubov bosons through
\begin{align}
\left(\begin{array}{c}
a_{k,1}\pm a_{-k,1}^{\dagger}\\
a_{k,2} \pm a_{-k,2}^{\dagger}
\end{array}\right) &  = \left(\begin{array}{cc}
u_k & v_{k} \\
v_{k} & -u_k
\end{array}\right)^{(*)}
\left(\begin{array}{c}
\gamma_{k} \pm \gamma_{-k}^{\dagger}\\
\beta_{k} \pm \beta_{-k}^{\dagger}
\end{array}\right),
\end{align}
with
\begin{eqnarray}
\!\!\!\!\!\!\!  u_k = \frac{i\sin(\frac{k}{4})+\cos(\frac{k}{4})}{\sqrt{2\cos(\frac{k}{2})}},
\;\;\;\;
\!\!\!\!\!\!\!  v_k = \frac{i\sin(\frac{k}{4})-\cos(\frac{k}{4})}{\sqrt{2\cos(\frac{k}{2})}},
\end{eqnarray}
where the conjugation ($*$) of the transformation matrix is taken for the ``$-$'' sign.
After this substitution and taking the limit of $m\rightarrow\infty$, the on-site correlator is given
by:
\begin{equation}
\tilde{S}_{0}^{xx}(\omega)   = \pi S \frac{\bar{\rho}\left ( \frac{\omega}{KS} \right)}{\omega}+\frac{\pi S}{2}\delta(\omega)\int_{-\pi}^{\pi}\frac{dk}{\cos(\frac{k}{2})},
\end{equation}
where the dimensionless density of states, $\bar{\rho}(\omega/(KS))=K S \rho(\omega)$, is defined as follows:
\begin{equation}
\bar{\rho}(x)=\frac{2}{\sqrt{1-\left(\frac{x}{2}\right)^{2}}}.
\end{equation}
The divergence at $\omega=2KS$ arises from  the Van Hove singularity in the density of single-magnon states at the top of the spin-wave band. 
Going back to the original reference frame, we have
\begin{equation}
S_{0}(\omega) =  \langle \tilde{S}_{0}^{xx}(\omega) \rangle,
\end{equation}
where we do need not to specify the superscript because ${S}_{0}^{xx}(\omega) = {S}_{0}^{yy}(\omega)={S}_{0}^{zz}(\omega)$.

\subsection{Nearest-neighbor dynamical structure factor}

There are two different contributions to the dynamical spin correlator between nearest-neighbor sites because of the two-site unit cell.
Let us first consider the $xx$ bond $(i,1)-(i,2)$.
The conservation of the flux operators $W_p$ implies that only the
correlator between the twisted $x$ spin components, $\langle\tilde{S}_{i,1}^{x}\tilde{S}_{i,2}^{x}\rangle$,
is non-zero on this bond. From the spin wave theory, we have
\begin{align}
\tilde{S}_{1}^{xx}(\omega)  \simeq \frac{1}{m}\sum_{k}e^{-ik/2} \langle \tilde{\chi}_{12}^{xx}(k,\omega) \rangle,
\end{align}
with
\begin{eqnarray*}
\langle \tilde{\chi}_{12}^{xx}(k,\omega) \rangle = {\pi^{2} \over \rvert\cos(\frac{k}{2})\rvert } \left[ e^{-i\frac{k}{2}} \delta(\omega-E_{k})-e^{i\frac{k}{2}} \delta(\omega)\right].
\end{eqnarray*}
In the  $m\rightarrow\infty$ limit, there is
\begin{equation}
\tilde{S}_{1}^{xx}(\omega) = \left[\frac{\pi \omega S }{2(KS)^{2}}-\frac{\pi S}{\omega}\right]\bar{\rho} \left (\frac{\omega}{KS} \right) 
-\frac{\pi S}{2}\delta(\omega) \!\! \int_{-\pi}^{\pi}\frac{dk}{\cos(\frac{k}{2})}.
\nonumber
\end{equation}

Similarly, only the $y$-components of the twisted spins, $\tilde{S}_{1}^{yy}(\omega)=\langle\tilde{S}_{i,2}^{y}\tilde{S}_{i+1,1}^{y}\rangle$, contribute to the NN spin correlator on the other $yy$ bond $(i,2)-(i+1,1)$. By  symmetry, this correlator is the same as the $xx$ correlator on the bond $(i,1)-(i,2)$ calculated above. Consequently, we can ignore the superscripts $xx/yy$ when referring to the NN spin correlator. Back to the original spin reference frame, we NN dynamic structure factor becomes
\begin{eqnarray}
S_{1}(\omega) =  \tilde{S}_{1}^{xx}(\omega).
\end{eqnarray}

\bibliographystyle{apsrev4-1}
\bibliography{paper}

\end{document}